\documentclass[longbibliography, aps, prx, amsmath, amssymb, amsfonts, twocolumn,superscriptaddress, footinbib]{revtex4-1}
\usepackage[german,american,english]{babel}
\usepackage{graphicx}
\usepackage{graphics}
\usepackage{dcolumn}
\usepackage{bm}
\usepackage{amssymb}
\usepackage{amsmath}
\usepackage{amsfonts}
\usepackage{epsfig}
\usepackage{mathbbol}
\usepackage{latexsym}
\usepackage{dcolumn}
\usepackage{subfigure}
\usepackage{hyperref}
\usepackage{float}
\usepackage[normalem]{ulem}
\usepackage[usenames, dvipsnames]{color}
\usepackage{epstopdf}

\hypersetup{                    
    colorlinks,
    citecolor=blue,
    filecolor=blue,
    linkcolor=blue,
    urlcolor=blue
}
\usepackage[utf8]{inputenc}

\begin{document}
\title{Speedup of the Quantum Adiabatic Algorithm using Delocalization Catalysis}


\author{Chenfeng Cao}
\affiliation{Department of Physics, The Hong Kong University of Science and Technology,\\ Clear Water Bay, Kowloon, Hong Kong, China}

\author{Jian Xue}
\affiliation{Institute of Physics, Chinese Academy of Sciences, Beijing, China}

\author{Nic Shannon}
\affiliation{Theory of Quantum Matter Unit, 
Okinawa Institute of Science and Technology Graduate University, Onna-son, Okinawa 904-0412, Japan}

\author{Robert Joynt}
\affiliation{Department of Physics, University of Wisconsin–Madison, 1150 University Avenue, Madison, Wisconsin 53706, USA}
\affiliation{Kavli Institute for Theoretical Sciences, University of Chinese Academy of Sciences, Beijing, China}

\begin{abstract}
We propose a method to speed up the quantum adiabatic algorithm using catalysis by many-body delocalization. This is applied to random-field antiferromagnetic Ising spin models.  The algorithm is catalyzed in such a way that the evolution approximates a Heisenberg model in the middle of its course, and the model is in a delocalized phase. We show numerically that we can speed up the standard algorithm for finding the ground state of the random-field Ising model using this idea.  We also demonstrate that the speedup is due to gap amplification, even though the underlying model is not frustration-free. The crossover to speedup occurs at roughly the value of the interaction which is known to be the critical one for the delocalization transition. We also calculate the participation ratio and entanglement entropy as a function of time: their time dependencies indicate that the system is exploring more states and that they are more entangled than when there is no catalyst.  Together, all these pieces of evidence demonstrate that the speedup is related to delocalization.  Even though only relatively small systems can be investigated, the evidence suggests that the scaling of the method with system size is favorable. Our method is illustrated by experimental results from a small online IBM quantum computer, showing how to verify the method in future as such machines improve.  The cost of the catalytic method compared to the standard algorithm is only a constant factor.      
\end{abstract}
\date{\today}
\maketitle

\section{Introduction}\label{Introduction}

In quantum computing, optimization problems play a special role. This is simply because optimization is ubiquitous in all areas of human endeavor.  Any significant speedups due to quantum optimization algorithms are guaranteed wide application in computation.  The quantum adiabatic algorithm (QAA) is a leading candidate for such a speedup \cite{farhi1}. Adiabatic computation model was proven to be polynomially equivalent to the gate-based model \cite{aharonov2008adiabatic}. Multiple tantalizing results continue to make the QAA attractive \cite{mott2017solving, subacsi2019quantum}, but provable speedups remain elusive \cite{lidar}.  Small energy gaps that thwart adiabaticity are of course the main issue \cite{young}.    

One interesting approach to improve the QAA borrows the concept of catalysis from chemistry.  When two reactants meet, they may or may not combine to form the stable compound, that is, they may or may not find the global ground state.  This process may be compared to the evolution in the QAA, with the meeting of the reactants corresponding to a close avoided level crossing.  In chemistry, another species is added to the system.  This species allows the reactants to overcome whatever barrier prevents the reaction.  In the QAA, we add a catalytic term to the Hamiltonian that is absent at the beginning and end of the evolution, but plays the role of helper at the time of the avoided crossing. Unlike chemical catalysis, we do not actually enlarge the Hilbert space.  However, one may take a simplified view of the chemical process (as is typically done in textbooks) and regard it as chiefly modifying the original Hamiltonian in the appropriate subspace.  That is the analogy we pursue here, and we note that the catalysis terminology has become standard in the present context.  See for example, Sec. VII.E of Ref. \cite{lidar}.

The QAA catalysis idea was proposed rather early in the history of QAA under the rubric of simply changing the path in Hamiltonian space and it was shown to work for an artificially designed problem by Farhi \textit{et al.} \cite{farhi2}.  This possibility also features in the analysis of the universality of the QAA \cite{aharonov2008adiabatic}.   One can also use the possibility of adjusting the scheduling to reduce Majorana-Landau-Zener tunneling \cite{ashhab}. A more targeted form of QAA catalysis using a term that suppresses transitions was proposed by \"{O}zg\"{u}ler \textit {et al.} \cite {ozguler}. However, the speedups achieved were rather small, and improving that method seems to be computationally demanding. The question of the general utility of QAA catalysis remains open.  

In this paper, we take an approach based on condensed matter physics ideas.  We use spin models that have a transition from a many-body localized (MBL) phase to a many-body delocalized (MBD) phase.  The paradigmatic examples, and the one that are best understood \cite{huse, imbrie}, are one-dimensional spin models.  We use the random-field Ising model (RFIM) as the problem to be optimized and the one-dimensional Heisenberg model as its delocalized counterpart. The catalytic part of the Hamiltonian is designed so that the system approximates an MBD quantum model in the course of the evolution, while the end point Hamiltonians are the ones chosen in the usual QAA.  In the RFIM we have the additional features that there are natural energy gaps due to local spin singlet formation.  The ground state of the model in the low-disorder regime has a high degree of local entanglement and is far from the classical Ising-type models that map to interesting classical optimization problems.  Hence the RFIM will be the main focus of the paper.  It is of course a standard testbed for quantum optimization. 

MBL is generally thought to occur in all dimensions, \cite{gornyi,basko}, and has been observed experimentally in one and two dimensions \cite{bordia,choi}.  There may be some interesting differences between in local spectral functions between one dimension and higher dimensions \cite {nandkishore}.   We have also applied our ideas to the two-dimensional case.  The results are very similar to those for one dimension. One may turn the logic of this paper around and use the speedup as a diagnostic for a MBL-MBD transition, \textit{i.e.}, as a tool for investigating MBL in a given model.  Thus our results give some support to the idea that any differences in spectral functions between one and two dimensions do not affect the properties investigated in this paper, and the simple existence of a localization length is probably the only crucial requirement.  However, the system sizes used here are small, and detailed investigation as a function of the localization length is not possible.  

Putting the problem into a more general context of optimization, however, it is also possible that a suitable catalyst Hamiltonian may be found for other problems by choosing the catalyst such that the eigenstates of the total Hamiltonian change from MBL to MBD.  We give some further details of this point of view for the RFIM below.        

Optimization of the ground state energy in the RFIM may be viewed semiclassically as optimizing the positions of domain walls.  As the system evolves in the course of the QAA, domain walls move to keep the energy low.  A close avoided crossing is due to a long move.  The off-diagonal matrix element of the Hamiltonian between the two levels involved in the crossing is small because the Hamiltonian is local and any perturbation theory expression for this element occurs only at high order.  In the catalytic approach, this semiclassical picture breaks down - the domain walls themselves are now delocalized quantum objects because of the additional Heisenberg-like terms in the Hamiltonian.  We can then expect to find larger matrix elements for the wall motion, which will result in larger gaps in the energy spectrum.  It is evident that this picture is closely related to the thermalization property of MBD systems that distinguishes them from MBL systems.  

In one-dimensional Heisenberg models, there is a gap of topological origin for integer spin \cite{haldane}. This motivated us to do a comparison of spin-1/2 and spin-1 models to see if the presence of this gap would contribute to a speedup of the algorithm.

The simulations are all for the closed QAA.  We have not done any detailed analysis of our method in an open quantum system context. This would be interesting, as MBL (if defined by its thermalization properties) does not survive strong coupling to a bath \cite {nandkishore}. However, we have run it on the IBM online system with positive results that we also present.  The system size is too small to constitute real evidence for our conclusions, but it indicates a direction for developments of our ideas as quantum computers become bigger and less noisy.   

One physical defining feature of any kind of localization is that there is no correlation between energy eigenvalues for eigenstates that are spatially separated beyond a characteristic length.  This leads to level crossings that are not avoided.  Gap amplification in the QAA, though not in the MBL context, has been investigated by numerous authors. The situation at present is that gap amplification that preserves the form of the eigenstates is known to be possible in general for frustration-free optimization problems \cite {boixo1, boixo2} and not possible for certain specific frustrated models.  This is done by a modification of the Hamiltonian.  In our case, the gap amplification comes from a physically-motivated time-dependent Hamiltonian that changes the character of the instantaneous eigenstates in the course of the evolution. 

There are five pieces of evidence that we present to establish the connection of speedup in the present method to the MBL-MBD transition.  The first is that, as a function of the interaction strength, there is a crossover from no speedup to speedup that occurs at approximately the value that is known to be critical one for the MBL-MBD transition in the many-body model.  The second is that the time dependence of the speedup (to be defined below) follows the time dependence of the catalytic delocalizing term.  Third, we observed the same transition in the calculated gap amplification.  This is in line with the well-known crossover from Poisson to Wigner-Dyson level statistics that accompanies the MBL-MBD transition, and which is evident even at rather small system sizes and near the ground state \cite{tsau}.    Fourth, we calculate the participation ratio, the classic diagnostic for localization, as a function of time, see that it also tracks the speedup, both in its time dependence, and in its dependence on the strength of the interaction. Finally, we compute the entanglement entropy as a function of time.  The overall magnitude and the time dependence indicates that the system is moving away from the hypersurface of separable states much more efficiently than it does in the absence of a catalyst, again evidence that MBD is involved in the speedup.   

In considering this evidence, it is important to keep in mind that there is no phase transition in the state of the quantum computer itself.  There is such a transition only in the ground state of an infinite system governed by a Heisenberg model with a random-field term, and the time-dependent Hamiltonian of our modified QAA approximates that only for a certain range of intermediate times.  The change in the state of the actual system is therefore a crossover, not a phase transition.     

A paper closely related to ours is that of Hormozi et al. \cite{Hormozi}.  These authors added random nonstoquastic terms to the Hamiltonian that turn on at the initial point and off at the final point of the evolution. The aim is similar: give the system more paths to overcome barriers and find the true ground state near the close avoided crossing.  However, the methods are opposite - ours is designed to reduce randomness, giving something close to an ordered model with known properties, in particular, large local gaps.  Ref. \cite{Hormozi}, in contrast, introduces more randomness into the problem.  Also, the additional terms in the Hamiltonian are not functions of the original Hamiltonian. In our method, once the original Hamiltonian is specified, the catalyst Hamiltonian is determined. Thus, while we average over the disorder in the original optimization problem to evaluate our method, we do not need to average over many additional terms.  We note also that the spin model investigated in Ref. \cite{Hormozi} is on a complete graph.  We do not know if our method is likely to be useful in such models, since we use features of quantum spins that are only evident when coordination numbers are low (in fact independent of system size).  We follow Ref. \cite{Hormozi} in that we judge the efficacy of our method by comparison to the results when the catalytic term is absent.  The connection of Anderson localization to QAA was first pointed out by Altshuler, Krovi, and Roland, who came to the quite pessimistic conclusion that it rendered the QAA ineffective \cite{altshuler}.  One way to get around this was proposed by Dickson \cite{dickson}.  It uses ancilla qubits.  The goal of the present work is similar to that of Ref. \cite{dickson} but the method is completely different. Two-particle catalysts have been proposed previously with the aim of providing another sort of non-stoquastic terms, and have been shown to have positive effects \cite{albash}. To our knowledge, none have made the connection to MBL.  Lychkovskiy has noted that problem Hamiltonians whose ground state is a product state but whose excited states are entangled might be particularly amenable to the QAA \cite{lychkovskiy} and our results perhaps lend some support to that idea.

From the standpoint of computer science, the method we present is entirely heuristic: we offer no speedup proofs.  However, it is not unprecedented that successful heuristic methods are later shown to offer certain provable improvements in efficiency. The simplex method in optimization theory is one example.

We describe our model and how it is simulated in Sec. \ref{Calculation Method} and give the details of our presentation methods in Sec. \ref{sec: presentation}.  We apply the method to spin-1/2 models in Sec. \ref{Results for Spin 1/2}, and to spin-1 models in Sec. \ref{Results for Spin 1}.  In Sec. \ref{Scaling} we investigate the efficacy of the method as the number of spins is increased. In Sec. \ref{Experiments}, we verified our method by an experiment on an IBM quantum device. The results are summarized and discussed in Sec. \ref{Conclusion}.   

\section{Calculation Method}\label{Calculation Method}

In the QAA method for this problem, we start a system of $N$ spins at $t = 0$ in the ground state of some simple Hamiltonian $H_{0}$ whose ground state is easy to prepare. This is $|\psi_{0}\rangle$.  In the absence of catalysis, the state evolves according to the time-dependent Hamiltonian $H_{qaa}$ which is

\begin{equation}
    H_{qaa} = f(t)H_{0} + g(t)H_{f}
\end{equation}
with, for example, $ f(t) = 1 - t/t_{a}$ and $ g(t) = t/t_{a}$. 

According to the adiabatic theorem \cite{born1928beweis}, if the minimum spectral gap $\delta_{m}$ is strictly greater than 0 and the evolution is slow enough, the final state $|\psi_{f}\rangle$ at $t=t_a$ will have a high fidelity to the ground state $|\psi_{gs} \rangle$ of $H_f$. 

Our modified method is as follows.  We add a catalytic term $H_{c}$ according to the recipe
\begin{equation}
    H_{qaa}(t) = f(t)H_{0} + g(t)H_{f} + h(t)H_{c}.
\end{equation}

Here $f(0) = g(t_a) = 1$ and 
$f(t_a) = g(0) = h(0) = h(t_a) = 0$.  

On quantum annealers, our method can be implemented directly. On gate-based quantum computers, e.g. IBM Q quantum processors, Suzuki-Trotter decomposition may be employed for the original and the catalyzed evolution. Clearly, the increase in cost, measured in the number of quantum gates applied, is a multiplicative constant.  In our simulation of such a gate-based machine, we use a fourth-order Runge-Kutta algorithm where the number of time steps is proportional to $t_a$. 

The problem Hamiltonian we choose for our one-dimensional  work is the nearest-neighbor RFIM on a ring of N spins:
\begin{equation}
    H_{f}=\sum_{k=1}^{N} h_{k}S_{z}^{k} + J \sum_{k=1}^{N}S_{z}^{k} S_{z}^{k+1}
\end{equation}
where $h_k$ are chosen uniformly from the interval $[-1, 1]$, The catalyst term is
\begin{equation}
    H_{c}=J \sum_{k=1}^{N} (S_{x}^{k} S_{x}^{k+1} + S_{y}^{k} S_{y}^{k+1}).
\end{equation}
The initial Hamiltonian represents a staggered field: 
\begin{equation}
    H_{0}= \sum_{k=1}^{N} (-1)^{k} S_{x}^{k}.
\end{equation}
It is very important to choose $H_0$ carefully. If $H_0$ is not staggered, it commutes with the Heisenberg Hamiltonian, thus creates symmetry-induced level crossings and associated small gaps.

In this paper, we set $\hbar=1$ and measure all energies in units of the width of the probability distribution for the random fields, and times in the inverse of this quantity.   

We use the evolution function
\begin{equation}
    h(t) = 2 \frac{t}{t_{a}} (1 - \frac{t}{t_{a}}).
\end{equation}
Thus when $t = \frac{1}{2}t_{a}$, the system Hamiltonian is the sum of a nearest-neighbor Heisenberg Hamiltonian and transverse field Hamiltonians. The Heisenberg Hamiltonian can be written as
\begin{multline}
H _ { H } = J \sum _ { k } S _ { z } ^ { k } \otimes S _ { z } ^ { k + 1  }\\ +
\frac { 1 } { 2 } J \left( \sum _ { k } S _ { + } ^ { k } \otimes S _ { - } ^ { k + 1  } + \sum _ { k } S _ { - } ^ { k } \otimes S _ { + } ^ { k + 1  } \right).
\end{multline}
The second term can interchange neighboring spins.  This moves domain walls and provides the system with paths toward the ground state.  

The two-dimensional model we use is just the obvious generalization of this and the same remarks concerning domain walls apply, though of course the walls are now one-dimensional objects, not points. 

It is known that the Heisenberg model in a random field undergoes a many-body localization at about $J\approx 1/3$ \cite{pal, luitz, xu} with the XX and YY terms driving the delocalization.  Thus during the evolution, the system is in a delocalized phase for $J$ sufficiently large and $t/t_a$ sufficiently close to $1/2$.  Of course for the finite, and indeed rather small, systems considered here there are no sharp transitions, and we will be looking for smooth transitions or crossovers.

In Secs. \ref{Results for Spin 1/2} and \ref{Results for Spin 1} we respectively simulate the evolution for spin-1/2 and spin-1 quantum systems. The ground state is calculated by exact diagonalization, and the time-dependent Schrödinger equation is solved numerically as described above. Each point in Sec. \ref{Results for Spin 1/2} and Sec. \ref{Results for Spin 1} is averaged over 32 realizations and each point in Sec. \ref{Scaling} is averaged over 256 realizations.

\section{Presentation Details}
\label{sec: presentation}
To understand what is happening in our simulations we will be computing various quantities that are now defined.   

We denote $P_{gs}(t_a)$ as the overlap between the final state $|\psi_{f} \rangle$ and the ground state $|\psi_{gs} \rangle$ of the final Hamiltonian: $P_{gs} = |\langle \psi_f | \psi_{gs} \rangle|^{2}$ at time $t_a$.  $\overline{P_{gs}}(t_a)$ is this quantity averaged over realization.  This has been a common measure of the quality of an approximate wavefunction in quantum many-body physics since the 1980s when exact diagonalization became possible for systems of small but non-trivial size.  See, \textit{e.g.,} Ref. \cite{he} for an example.  $\overline{P_{gs}}(t_a)$ is also the average fidelity.  In more concrete terms, it is the average chance of getting the right answer when a measurement is made.

In considering the merit of the algorithm, it is very important to keep in mind that we do not need $ 1 - P_{gs} << 1$ for success.  Measuring the system repeatedly in the computational basis and evaluating the resulting energy each time, we achieve the proper result with probability $1 - (1-P_{gs})^R$ after $R$ repetitions.  If, for example, $P_{gs} = 1/2$, then 10 repetitions will give the ground state with a probability greater than 0.999.      

There are several advantages to plotting $\overline{P_{gs}}(t_a)$.  If one is interested in the relative speed of two algorithms, then once one has decided on the success probability that defines the conclusion of the algorithm, one can simply take a horizontal cut through the graph and read off the ratio of the abscissas of the two points thus determined.  We will not do that, since we are more interested in the relative accuracy of the algorithm for a fixed run time, so we take a vertical cut.  The average overlap of the computed ground state and the actual ground state with catalysis is defined as $P^c$, and the average overlap without catalysis as $P^0$.  Then the speedup $SP$ is the ratio of $P^{c}$ and $P^{0}$, $SP = P^{c}/P^{0}$, which will be plotted below as a function of the interaction strength and the time.  

Thus the $\overline{P_{gs}}(t_a)$ plots give more information than simply tabulating, for example, run times.  In fact, they give a very nuanced picture of the ever-present tradeoff between accuracy and run time in difficult computational problems.  A further, and for us crucial, reason we chose this presentation method is that for our particular algorithm, in which the catalyst Hamiltonian is turned on and then off, we can identify signatures of the success or failure of the idea of catalysis in the details of the time dependence, as we shall see below. 

In adiabatic quantum computing, the required annealing time of QAA for a fixed error size is given by an expression of the form

\begin{equation}
T = O \left( \frac { \left \| \frac{d}{dt} \tilde{H}(t) \right \| } { \delta_{m} ^ { 2 } } \right).
\end{equation}

It is of great interest to check that the positive effects of $H_c$ do arise from gap amplification.  Hence we will present the minimum spectral gap $\delta_m$ with and without the catalytic term. $\delta_{m}$ is the difference between the ground state energy and the first excited state energy minimized over all times $0 \leq t \leq t_a$.  Small gaps arise when low-energy states have small overlaps, and the latter is a direct consequence of localization.  Hence the minimum gap should be smaller when we are in the MBL phase, and plots of $\delta_m$ are direct way to illustrate that.  

An important quantity related to delocalization is the inverse participation ratio $IPR$, defined for spin 1/2 as
\begin{equation}
IPR(t) = \sum_{\alpha =0}^{2^N-1} | \langle \alpha | \psi (t) \rangle |^4,
\end{equation}
where $| \alpha \rangle $ runs over the computational basis.  If $IPR=1$, then $|\psi \rangle$ is completely localized in this basis, and delocalized when $IPR << 1$.
$\overline{IPR}$ is this quantity averaged over realizations of the disorder.  This quantity is the canonical measure of localization.  It should be recognized, of course that the participation ratio is basis-dependent.  In ordinary single-particle quantum mechanics the natural basis is the position basis.
Here the computational basis plays that role.  In this basis the initial state is completely delocalized with a very small $IPR$ and the final state, if properly found, is completely localized with $IPR=1$.  The chief way for a computation using the QAA to fail is \textit{premature} localization - the system gets stuck in the wrong state and its overlap with (also localized) state is small.  The catalyst term is designed to prevent this.  Hence we would hope to see the $IPR$ rise more slowly when the catalyst is added.  If this is the case, then MBD is at work.   

The last diagnostic for the system is the entanglement entropy as a function of time.  We divide the system randomly into equally sized two parts A/B and calculate the reduced density matrix of A and the bipartite entanglement entropy according to standard prescriptions:
\begin{equation}
    \rho_{A}(t) = \operatorname{Tr}_{B}|\psi (t)\rangle \langle \psi (t)|
\end{equation}
\begin{equation}
    \mathcal{S}(\rho_{A}(t))=-\operatorname{Tr}\left[\rho_{A}(t) \log \rho_{A}(t)\right].
\end{equation}
If $\mathcal{S}(\rho_{A}(t))$ is large, then we expect that quantum information can pass more freely throughout the system.  This is clearly desirable in a search for a global minimum of any non-local operator in the Hilbert space.  $\mathcal{S}(\rho_{A}(t))=0$ at the initial and final times, the first true by construction, the second by the nature of the solution state.  It peaks in the middle of the evolution.  If MBD is behind the observed speedup, then we would expect to see a much larger peak in $\mathcal{S}(\rho_{A}(t))$ for the catalyzed evolution.  The peak should occur at later times as well, since premature localization will tend to suppress $\mathcal{S}(\rho_{A}(t))$. 

\section{Results for Spin 1/2}\label{Results for Spin 1/2}

The ground state of the spin-1/2  nearest-neighbor antiferromagnetic Heisenberg chain is not ordered. Instead it is quantum-critical with power-law spin correlations.  Due to the low dimensionality, neighboring spins have very strong singlet correlations, and therefore a large amount of entanglement at a local level. Local formation of triplet pairs costs a large energy.   Nevertheless, the model does not have a gap due to the Lieb-Schultz-Mattis theorem \cite{lieb}.  The low-energy excitations are spinon pairs on top of a background of resonating valence bonds.  The spinons have a large short-range repulsion but are otherwise deconfined.  

We now use the Heisenberg antiferromagnetic model to catalyze the evolution from a 1-dimensional antiferromagnet to a 1-dimensional random-field Ising spin chain.
We present our results using several methods, in order to highlight both the sheer improvement that one can make to the QAA by introducing the delocalization catalytic term , and to make clear the connection to the MBL-MBD transition. 

The results for spin 1/2 are shown in Fig. 1, Fig. 2 and Fig.3(a) for a system of $N=12$ spins.  

\begin{figure}[h!]
\centering
\subfigure[]{
\label{fig:subfig:a} 
\includegraphics[scale=0.45]{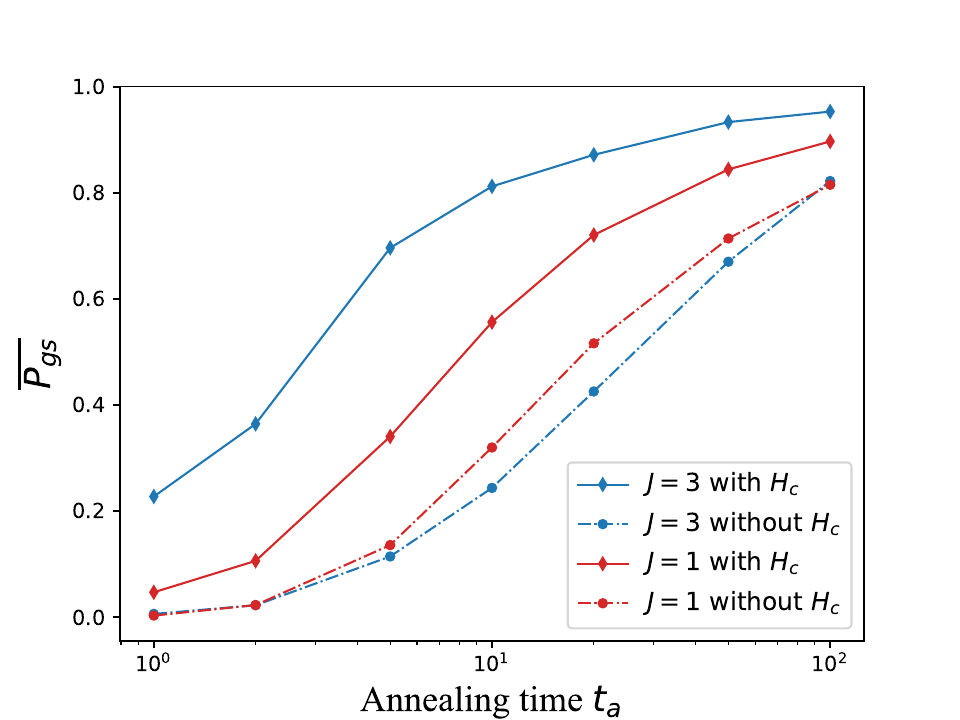}
}
\subfigure[]{
\label{fig:subfig:a} 
\includegraphics[scale=0.45]{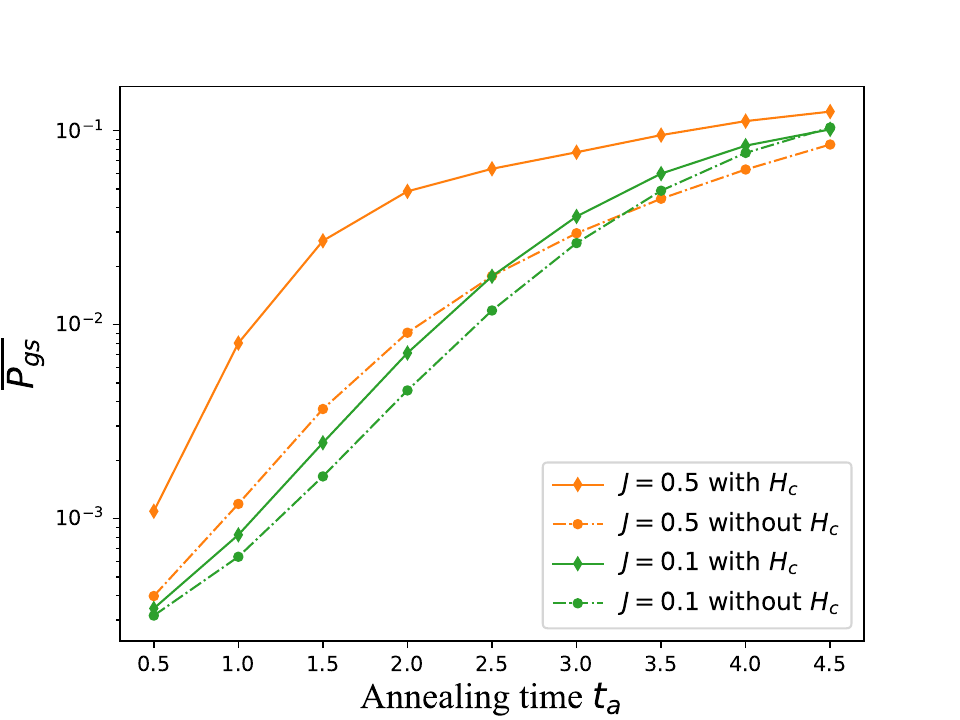}
}
\caption{Average ground state probabilities and minimum gaps for spin-$1/2$ quantum chain with 12 spins. Overlap of the ground state and the final wavefunctions calculated by the catalyzed (solid lines) and uncatalyzed (dashed lines) QAA. Shown in (a) are the results for $J = 1,  3$ and in (b) for $J= 0.1, 0.5$. Catalysis speeds up the algorithm substantially, particularly for $J=3$, but hardly at all for $J=0.1$.}
\label{fig:1}
\end{figure}

\begin{figure}[h!]
\centering
\subfigure[]{
\label{fig:subfig:a} 
\includegraphics[scale=0.45]{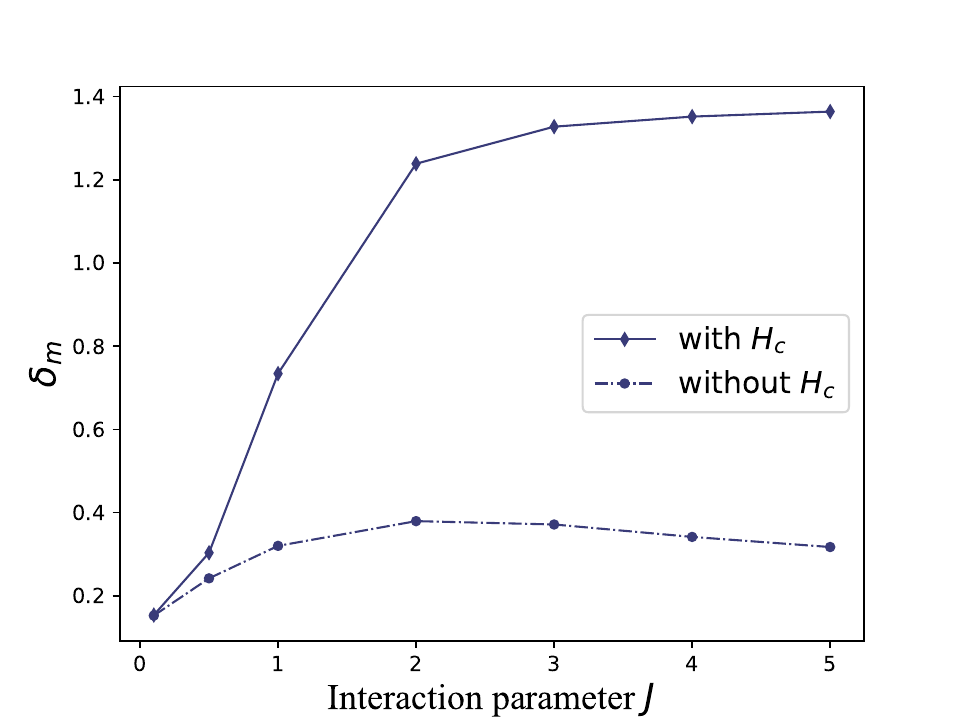}
}
\subfigure[]{
\label{fig:subfig:a} 
\includegraphics[scale=0.45]{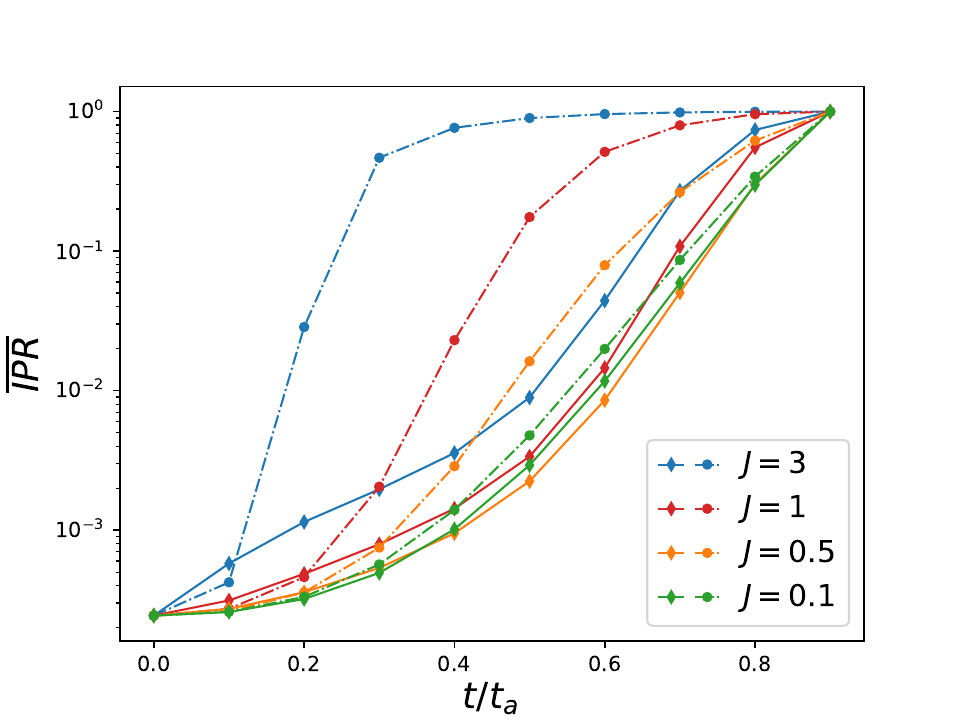}
}
\subfigure[]{
\label{fig:subfig:a} 
\includegraphics[scale=0.45]{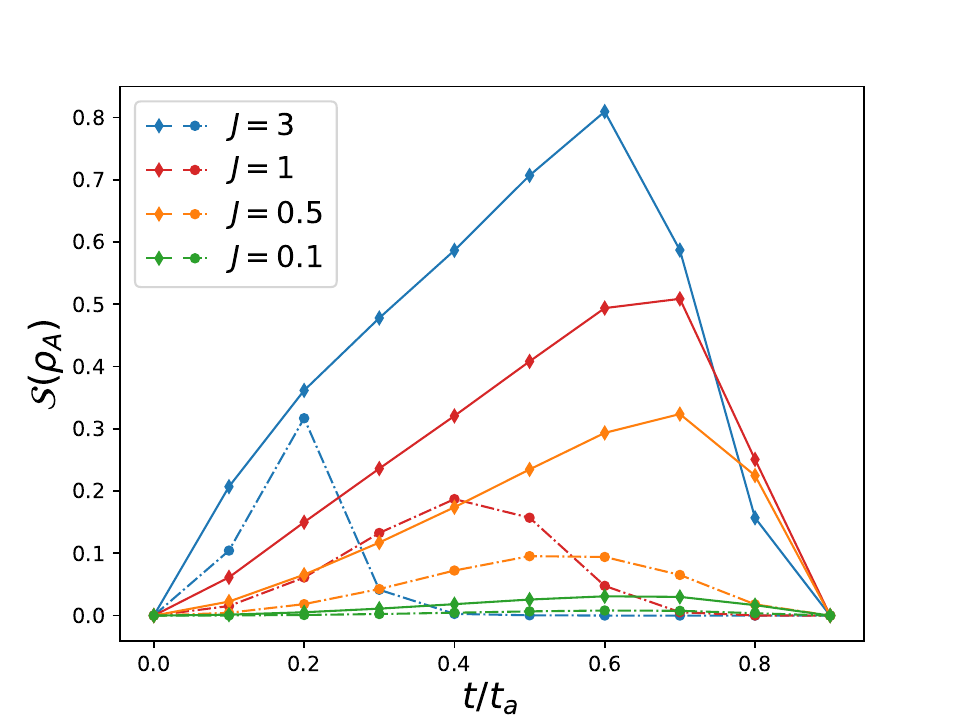}
}
\caption{Minimum gaps, inverse participation ratios and entanglement entropy for spin-$1/2$ quantum chain with 12 spins. (a) Average minimum gap with or without catalysis vs. interaction parameter $J$. (b) Average inverse participation ratio for catalyzed (solid curves) and uncatalyzed (dashed curves) cases. (c) Average entanglement entropy for catalyzed (solid curves) and uncatalyzed (dashed curves) cases.}
\label{fig:1}
\end{figure}

In Fig. 1(a) the interaction between neighboring spins is relatively large (i.e. $J \geq 1$).  For this case the optimization process can be accurately thought of as the appropriate motion of domain walls, precisely the physical process that the catalyst Hamiltonian $H_{c}$ is designed to promote.  We see that at $J=3$ the catalysis increases the overlap of the computed ground state and the actual ground state dramatically for any evolution time and the increase is also substantial when $J=1$.  In Fig. 1(b) we see that the crossover for speedup to no speedup occurs as we go from $J=0.1$ to $J=0.5$, which is where the MBL-MBD transition takes place in the bulk system.  In Fig. 2(a) we plot $\delta_m$ against $J$ with and without $H_c$.  The crossover as a function of $J$ is evident.  The plots for the two algorithms are very differnt when $J$ exceeds the critical value:  $J\geq 1/2$, but are more or less the same when  $J\leq 1/2$.  This very considerable amplification of the minimum gap is closely associated with the speedup of the algorithm, as comparison with Fig. 1(a) shows. 

To provide the most direct possible evidence for connection between MBL and speedup, we plot $IPR$ in Fig. 2(b).  Again, the curves for the two algorithms are nearly the same for $J=0.1$ but begin to differ substantially at $J=0.5$ and the continue to be very different for larger $J$ values. The quick upturn in the uncatalyzed case is a sign or premature localization. This problem is greatly reduced in the catalyzed algorithm, as expected.  The results for the entanglement entropy are shown in Fig. 2(c).  The peak for the catalyzed case occurs at roughly the same point in time as the upturn in the $IPR$, confirming that they come from the same cause. Even more impressive, the the peak value becomes much higher with catalysis. This large enhancement is a sign that the Hilbert space is better explored - the path in Hilbert space is straying much further away from the manifold of product states.

\begin{figure}[h!]
\centering
\subfigure[]{
\label{fig:subfig:a} 
\includegraphics[scale=0.45]{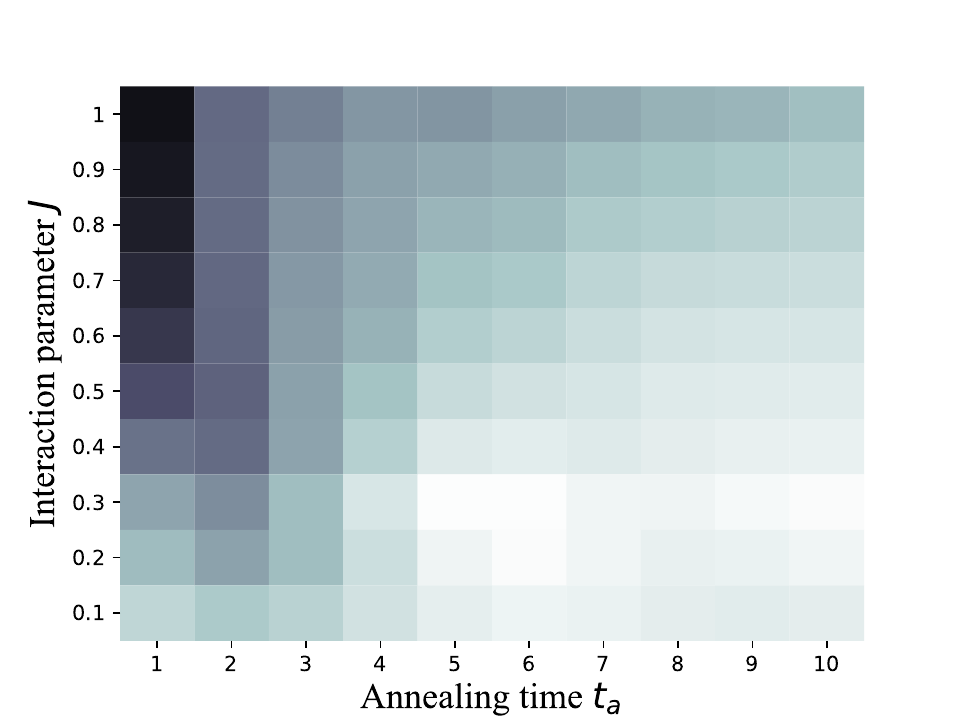}
}
\subfigure[]{
\label{fig:subfig:a} 
\includegraphics[scale=0.45]{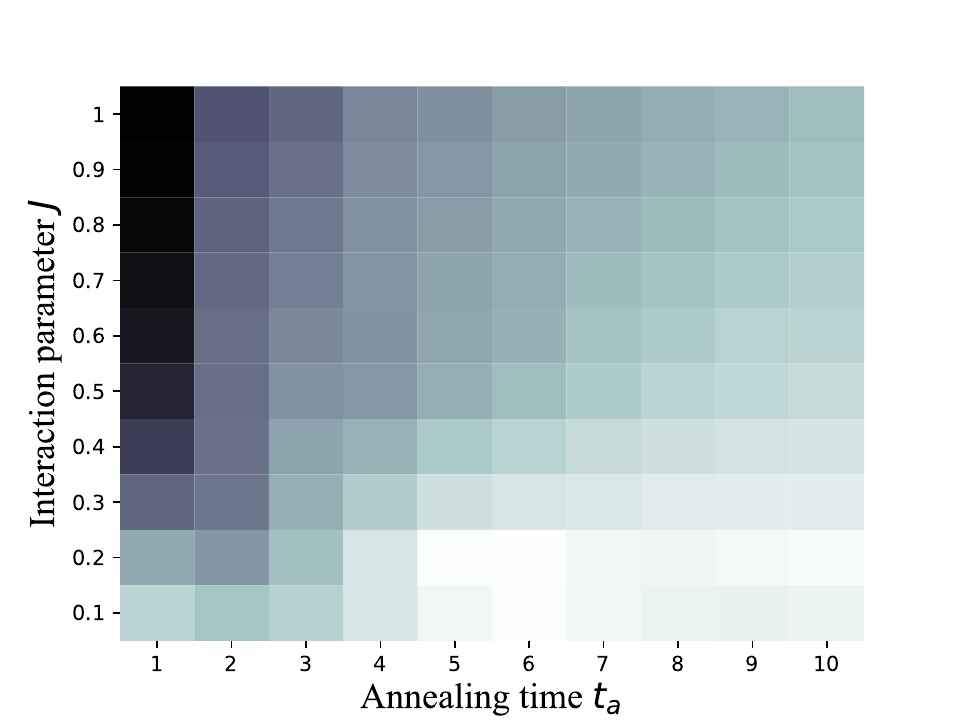}
}
\subfigure[]{
\label{fig:subfig:a} 
\includegraphics[scale=0.55]{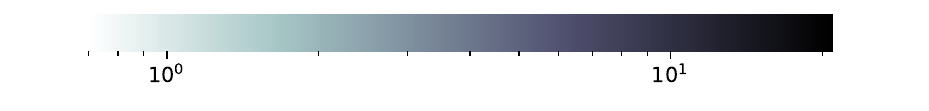}
}
\caption{Speedup SP as a function of $J$ and $t_a$. Darker regions correspond to higher speedups. (a) Speedup for spin-1/2 quantum chain with 12 spins. (b) Speedup for 4 x 3 spin-1/2 grid. (c) The color bar shows the speedup values. }
\label{fig:1}
\end{figure}

Of particular note from the optimization point of view is that the method actually works well in the region of moderate coupling.  This is where the optimization problem is the most difficult.  Naive methods work well both for $J<<1/2$ and $J>>1/2$. 

From the MBL point of view, the important fact is that the transition from no speedup to speedup around $J=1/2$ matches approximately the MBL-MBD transition.  The latter takes place in the static model at $J\approx 1/3$, which is roughly the average value of $J$ in the course of the QAA evolution.  This justifies the idea that speedup can serve as a signal of the many-body transition.


The one-dimensional RFIM is a fairly simple model. We can directly extend the whole method to the two-dimensional square lattice.  The initial transverse magnetic field is again staggered in the x-direction.  The ground state of the two-dimensional nearest-neighbor Heisenberg model that catalyzes the evolution model is quite different from the one-dimensional version.  It is an ordered magnet though with an ordered moment that is substantially reduced by quantum effects \cite{chakravarty},  spins have a much more classical character. Nevertheless, the concept of domain wall motion remains important, we expect the domains to be delocalized in the MBD phase as they are in one dimension.  Indeed, motion of density domain walls in a two-dimensional disordered boson system has been observed \cite{choi}.

 Fig. 3(b) shows the speedup map for the 4 x 3 spin-1/2 grid. The effect of the 2-dimensional catalyst Hamiltonian is very similar to the 1-dimensional case. For small $J$, the lack of speedup is analogous to what happens in one dimension, while for larger $J$, the speed up is obvious since the minimum gap can be efficiently amplified.  There do not appear to be any qualitative differences between 1 and 2 dimensions, and even quantitative differences are very minor.

\section{Results for Spin 1}\label{Results for Spin 1}

The ground state of the spin-1 antiferromagnetic Heisenberg chain is also not ordered, but its spin correlations are exponential, not power-law.  It exhibits the Haldane gap \cite{haldane}.  One may think of the ground state as being effectively spontaneously dimerized, as shown by the Affleck-Kennedy-Lieb-Tasaki (AKLT) construction \cite{affleck} for the ground state of a closely related Hamiltonian also to be used below.  The low-energy excitations are massive spin waves.  However, disorder can also induce localized fractional (spin-1/2) excitations, which also appear at the boundaries if open boundary conditions are employed.  

\begin{figure}[h!]
\centering
\subfigure[]{
\label{fig:subfig:a} 
\includegraphics[scale=0.45]{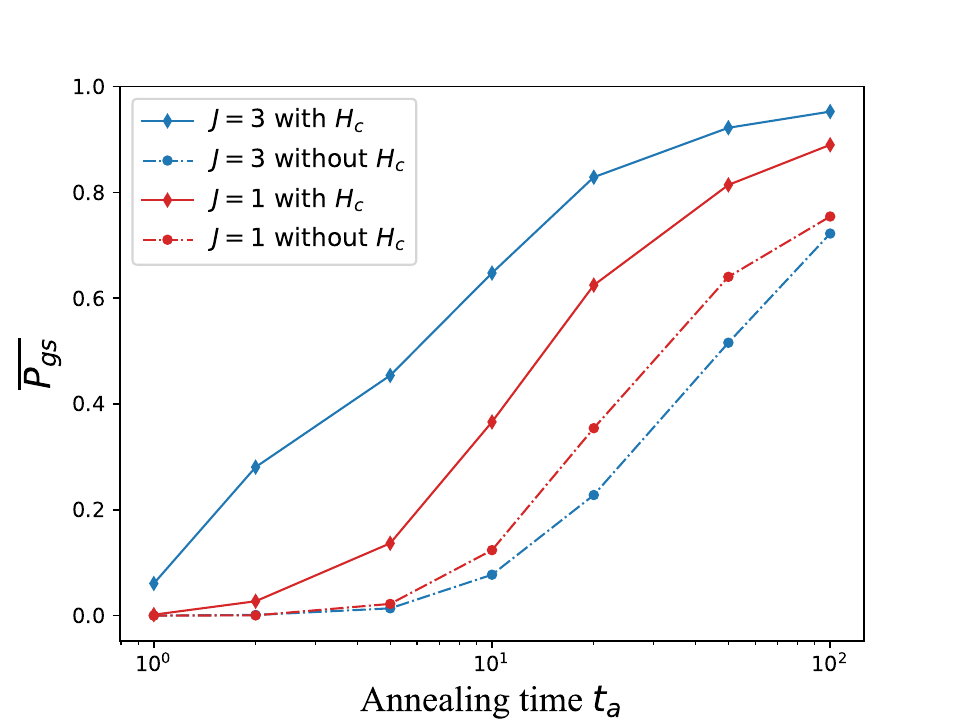}
}
\subfigure[]{
\label{fig:subfig:a} 
\includegraphics[scale=0.45]{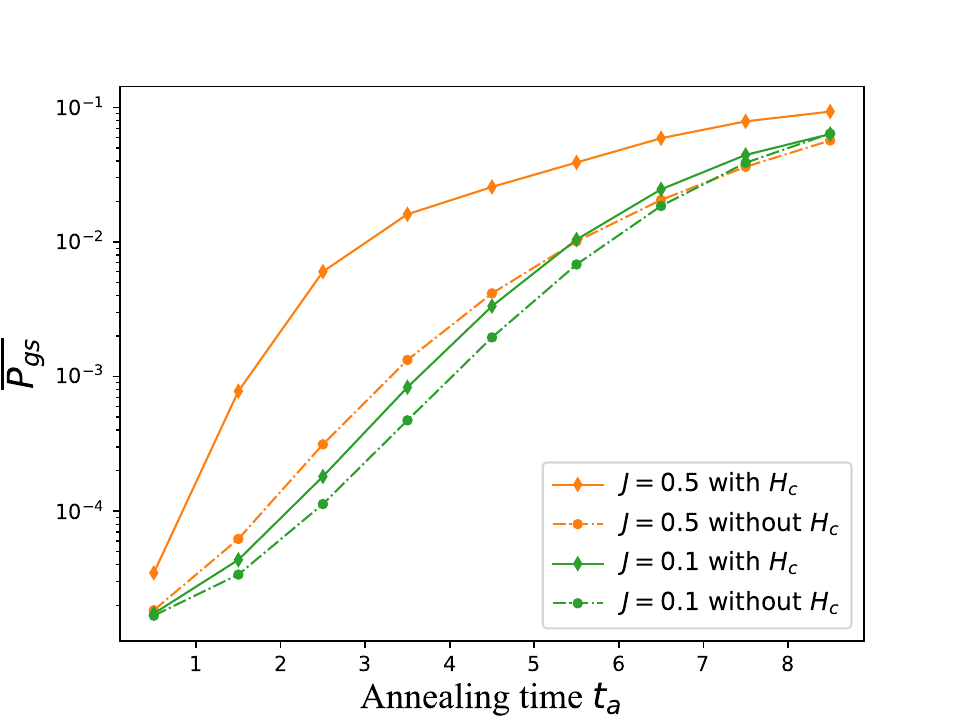}
}
\caption{Average ground state probabilities and minimum gaps for spin-$1$ quantum chain with 8 spins. Overlap of the ground state and the final wavefunctions calculated by the catalyzed (solid lines) and uncatalyzed (dashed lines) QAA. Shown in (a) are the results for $J = 1,  3$ and in (b) for $J= 0.1, 0.5$. Catalysis speeds up the algorithm substantially, particularly for $J=3$, but hardly at all for $J=0.1$.}
\label{fig:1}
\end{figure}

\begin{figure}[h!]
\centering
\subfigure[]{
\label{fig:subfig:a} 
\includegraphics[scale=0.45]{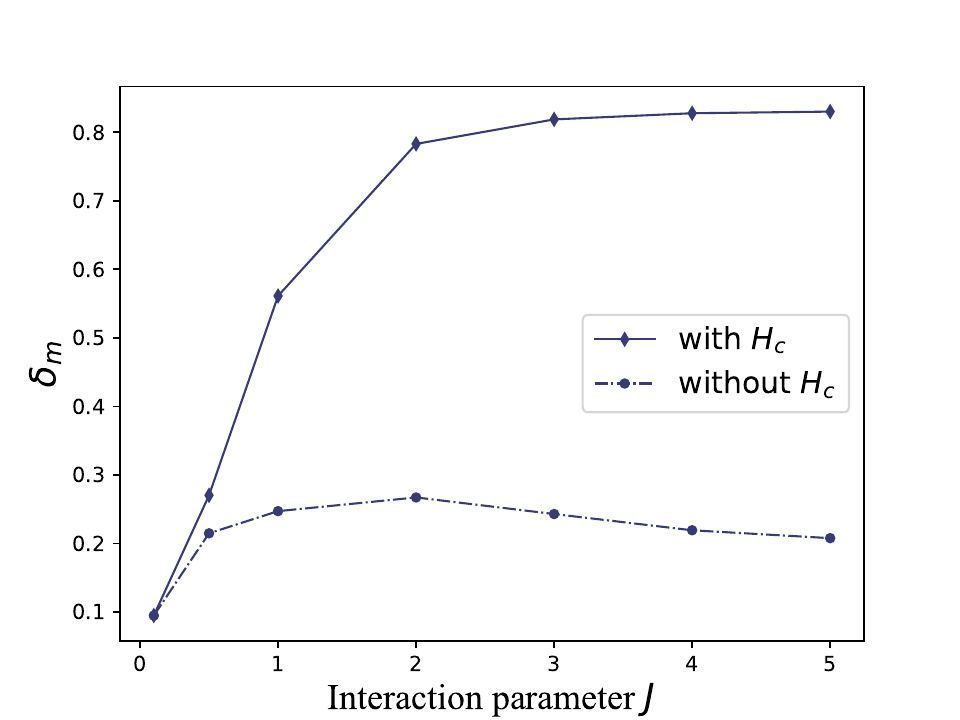}
}
\subfigure[]{
\label{fig:subfig:a} 
\includegraphics[scale=0.45]{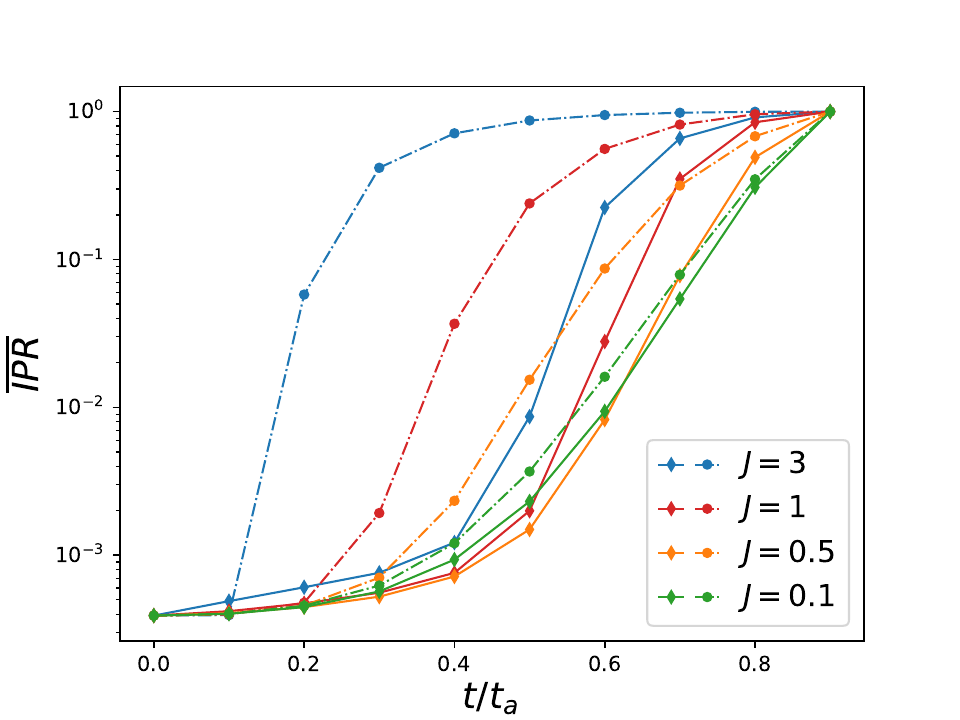}
}
\subfigure[]{
\label{fig:subfig:a} 
\includegraphics[scale=0.45]{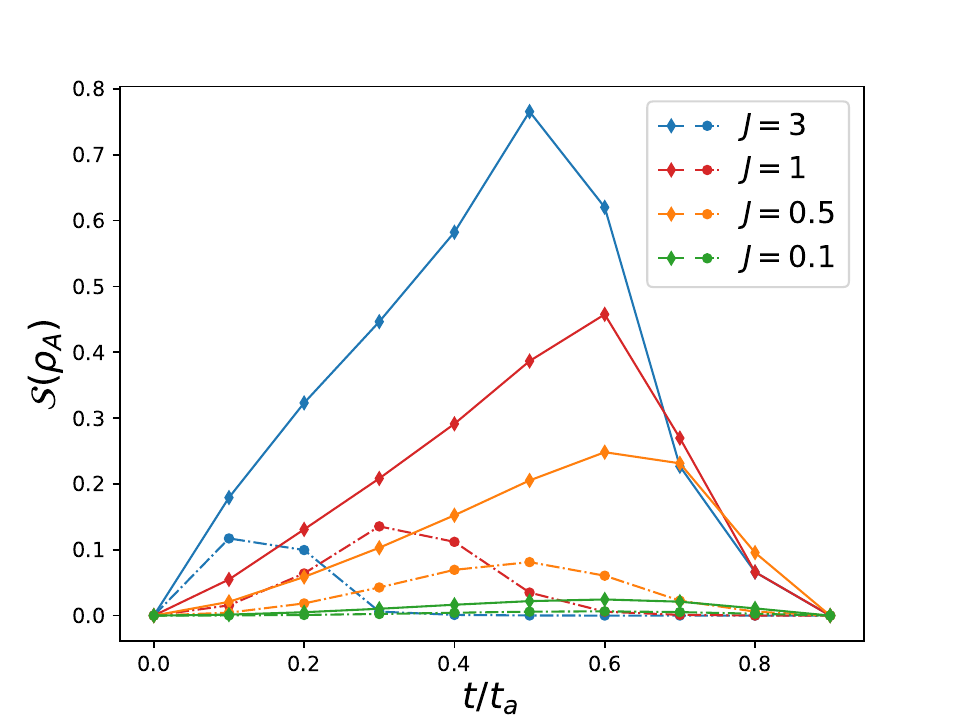}
}
\caption{Minimum gaps, inverse participation ratios and entanglement entropy for spin-$1$ quantum chain with 8 spins. (a) Average minimum gap with or without catalysis vs. interaction parameter $J$. (b) Average inverse participation ratio for catalyzed (solid curves) and uncatalyzed (dashed curves) cases. (c) Average entanglement entropy for catalyzed (solid curves) and uncatalyzed (dashed curves) cases.}
\label{fig:1}
\end{figure}

This is the second model that we use to catalyze the evolution from a 1-dimensional nearest-neighbor antiferromagnet to a 1-dimensional nearest-neighbor RFIM spin chain.  It is easy to show that finding the ground state of the spin 1 model also gives the solution to the problem of finding the RFIM ground state. Again the expectation is that local correlations will gap out close avoided level crossings.  We wish to test whether the existence of the Haldane gap will increase this effect or not.

MBL also occurs in the spin-1 chain \cite{chandran}, but it is important to note that the promotion to spin-1 from spin-1/2 also has several clear disadvantages.  The most obvious is that the computation time for fixed number of spins increases since the Hilbert space is bigger. Related to this is the fact that the energy level density is increased by a factor of $(3/2)^N$, where $N$ is the number of spins, with a corresponding decrease in the average separation between energy levels.  Finally, as the spin $S$ increases, the spins become more classical in the sense that the large energy difference between the singlet and triplet energies for a pair of spins becomes less evident with increasing $S$.  This is clearly not in line with what we believe to be the advantages of our method.

The numerical results for the spin-1 quantum chain are shown in Fig. 4(a) for $J=1$ and $J=3$.  The speedup is very similar to the spin-1/2 case. Again the catalyst Hamiltonian can speed up the evolution significantly.  In Fig. 4(b), we show that even for weak interactions there is some speedup as long as the evolution time is short.  Thus, as we saw in the spin-1/2 case, the size of the speedup increases as the interaction becomes stronger.  Gap amplification is present as well, as shown in Fig. 5(a).  It is slightly less strong than in the spin-1/2 case, reflecting the compression of energy levels. The initial spectral gap is an upper bound of the minimum gap $\delta_{m}$. For spin-$1/2$ quantum chain, the initial spectral gap is 2, but for spin-$1$ quantum chain, the initial gap is 1. The results of inverse participation ratio and entanglement entropy are  shown in Fig. 5(b)(c), and they are extremely similar to those in those in spin-$1/2$ case, so the same conclusions clearly apply.  We would only remark that at present we do not know how to disentangle any effects of the Haldane gap from the other effects we showed to exist in the spin-1/2 case.  Hence the effects of topology remain an open question.

\section{Scaling}\label{Scaling}

In this section, we investigate the scaling properties of the method.  We fix an annealing time $t_{a}$ and measure the speedup by $SP = P^{c}/P^{0}$, where $P^c$ is the average overlap of the computed ground state and the actual ground state with catalysis, $P^0$ is the average overlap without catalysis. 

For spin-1/2 quantum chain with different sizes, the speedups $SP$ for short annealing time $t_{a} = 1$ and long annealing time $t_{a} = 20$ are shown in Fig. 6(a)(b), as a function of the interaction strength $J$ and the number of spins $N$. The speedup for short-time annealing increases exponentially as $N$ increases in this range of parameters for all interaction strengths $J$. However, the speedup for long-time annealing is only scalable for large $J$. 

\begin{figure}[h!]
\centering
\subfigure[]{
\label{fig:subfig:a} 
\includegraphics[scale=0.45]{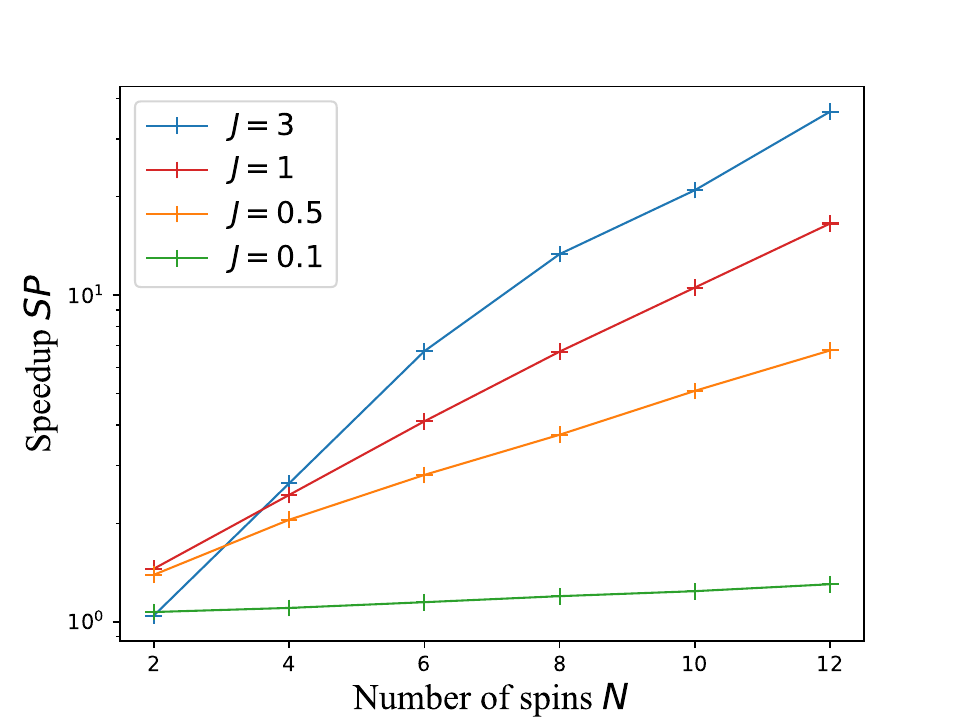}
}
\subfigure[]{
\label{fig:subfig:a} 
\includegraphics[scale=0.45]{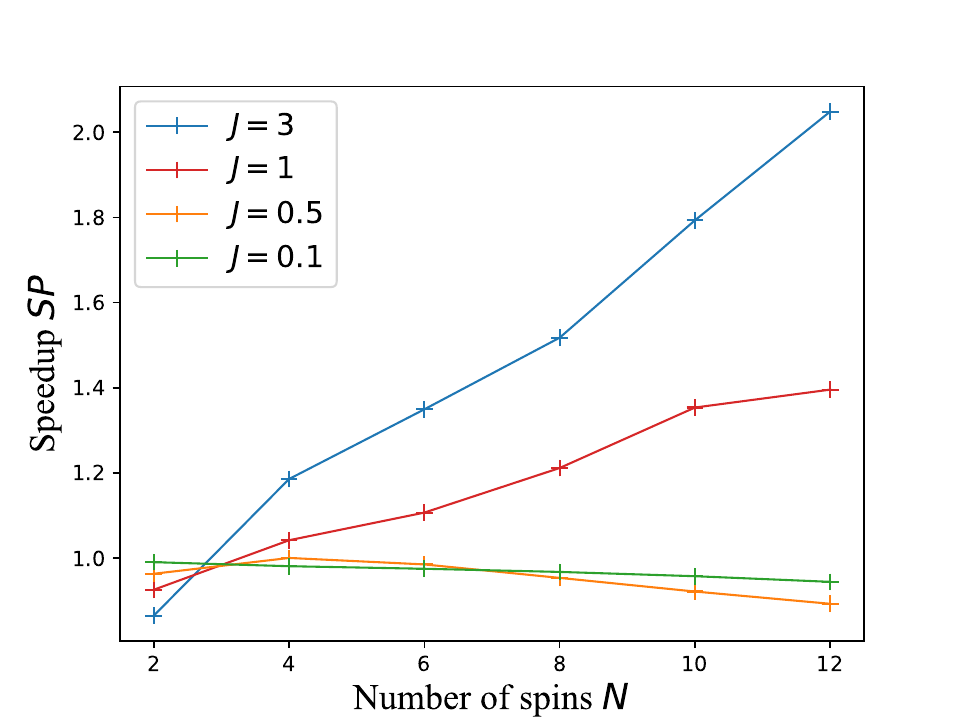}
}
\subfigure[]{
\label{fig:subfig:a} 
\includegraphics[scale=0.45]{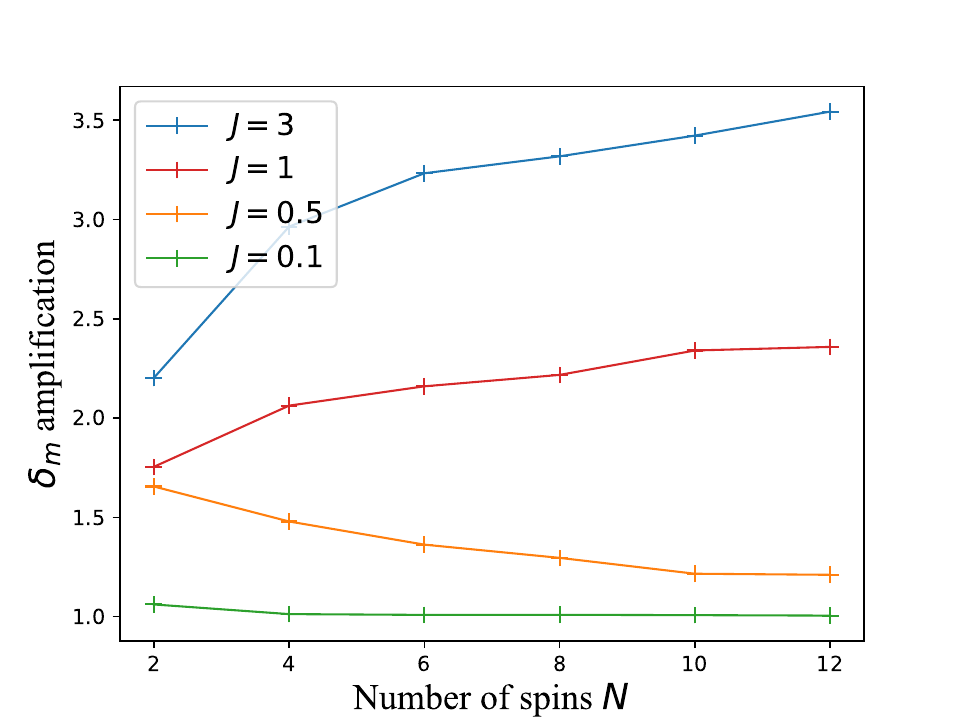}
}
\caption{Speedup vs. number of spins for spin-1/2 quantum chain with (a) $t_{a}=1$ (b) $t_{a}=20$. (c) Gap amplification vs. number of spins for spin-1/2 quantum chain with different interaction strengths $J$.}
\label{fig:3-1}
\end{figure}

In addition, we investigate the scaling effect of gap amplification, denote the minimum gap without catalysis as $\delta_{m}^0$, the minimum gap without catalysis as $\delta_{m}^c$, we measure the gap amplification by $\delta_{m}^c/\delta_{m}^0$. The result is shown in Fig. 6(c), which is consistent with the scaling of long-time annealing in Fig. 6(b). Therefore we conclude that the speedup for short-time annealing is not (solely) due to the gap amplification.

This analysis is limited to a small number of spins and a short evolution time, but there is a strong suggestion that the conclusions hold for larger systems: the positive results reported above are not an artifact of small $N$. Similar scaling phenomenon can be observed for spin-1 quantum chain.

\section{Experiments on an IBM Quantum Computer}\label{Experiments}

In this section, we run QAA algorithm on a real IBM quantum computer to test our method in the context of a real system with errors. The device we use is ibmq\_santiago \cite{ibmq_santiago}, which is a gate-based programmable quantum computer with 5 qubits. The structure of the quantum computer is shown in Fig. 7. 

\begin{figure}[h!]
\centering
\includegraphics[scale=0.5]{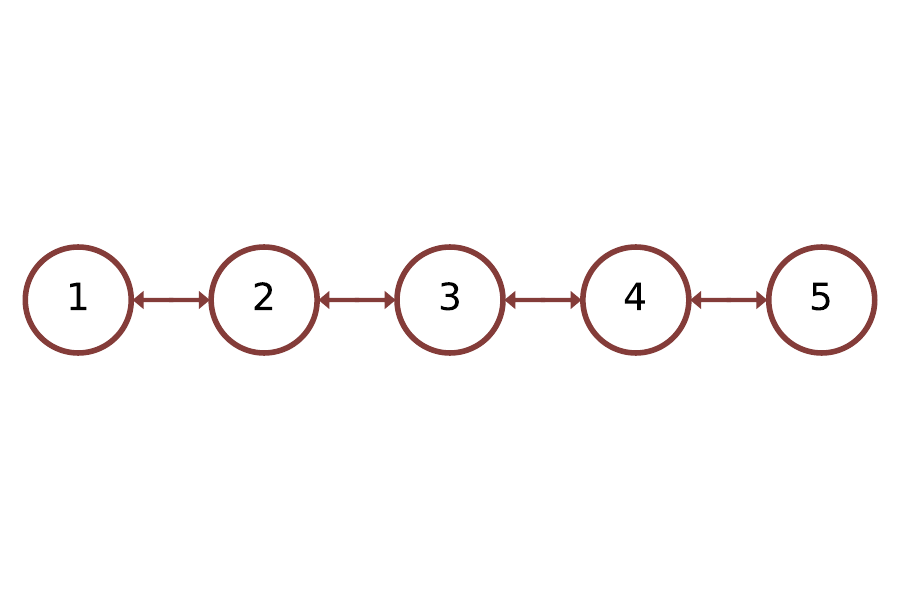}
\caption{Structure of ibmq\_santiago}
\label{fig:2}
\end{figure}

We run our circuits on a 5-qubit chain ordered by qubit-1, 2, 3, 4, 5. The staggered initial Hamiltonian is

\begin{equation}
    H_{0}= \sum_{k=1}^{5}(-1)^{k} \sigma_{x}^{k}.
\end{equation}

The final Hamiltonian is
\begin{equation}
    H_{f}= - \frac{3}{4} (\sigma_{z}^{1} + \sigma_{z}^{2} + \sigma_{z}^{4}) + J \sum_{k=1}^{4} \sigma_{z}^{k}\sigma_{z}^{k+1}.
\end{equation}

The distribution of field strengths is still random in the sense that there are fields only on qubits 1,2 and 4, but we do no averaging over realizations.

The catalyst Hamiltonian is 
\begin{equation}
    H_{c}= J \sum_{k=1}^{4} (\sigma_{x}^{k}\sigma_{x}^{k+1} +  \sigma_{y}^{k}\sigma_{y}^{k+1}).
\end{equation}

Since ibmq\_santiago is not a quantum annealer, we need to do Trotter decomposition to implement the evolution \cite{smith2019simulating}. First, we discretize the evolution time $t_{a}$ so that the evolution operator $U(t)$ becomes a product of discrete interval operators $\{U(\Delta t)\}$, then we approximate these operators with CNOT and single-qubit gates. 

Operators $e^{-i \sigma_{x} \Delta t}$ and $e^{-i \sigma_{z} \Delta t}$ can be implemented by single-qubit rotation gates $R_{x}$ and $R_{z}$ directly. Operator $N(\alpha, \beta, \gamma) = \exp \left(i\left(\alpha \sigma_{x} \otimes \sigma_{x}+\beta \sigma_{y} \otimes \sigma_{y}+\gamma \sigma_{z} \otimes \sigma_{z}\right)\right)$ can be decomposed by the circuit in Fig. 8, up to a global phase \cite{vatan2004optimal}. With Trotterization and gate decomposition, we can simulate a continuous time evolution with ibmq\_santiago.

\begin{figure}[h!]
\centering
\includegraphics[scale=0.8]{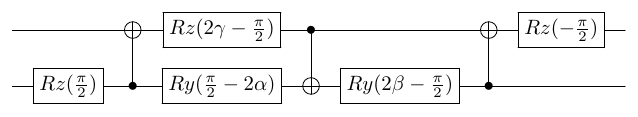}
\caption{Gate decomposition of $N(\alpha, \beta, \gamma)$}
\label{fig:2}
\end{figure}

The CNOT error rate of ibmq\_santiago is not negligible; it ranges from $5.054 \times 10^{-3}$ to $6.641 \times 10^{-3}$.  This means we can only simulate a short-time evolution. For annealing times $t_{a} = 0.2, 0.4, 0.6, 0.8, 1$, the probabilities of the ground state are shown in Fig. 9. The catalyst term can speed up the evolution efficiently for large interaction parameter $J$, which is consistent with the theoretical expectations and the earlier simulation results.  Of course the system size is too small to claim that the success seen here gives real evidence for our conclusions.  However, it does show how to implement the method on an actual computer, which we will be important as these machines improve.

\begin{figure}[h!]
\centering
\includegraphics[scale=0.45]{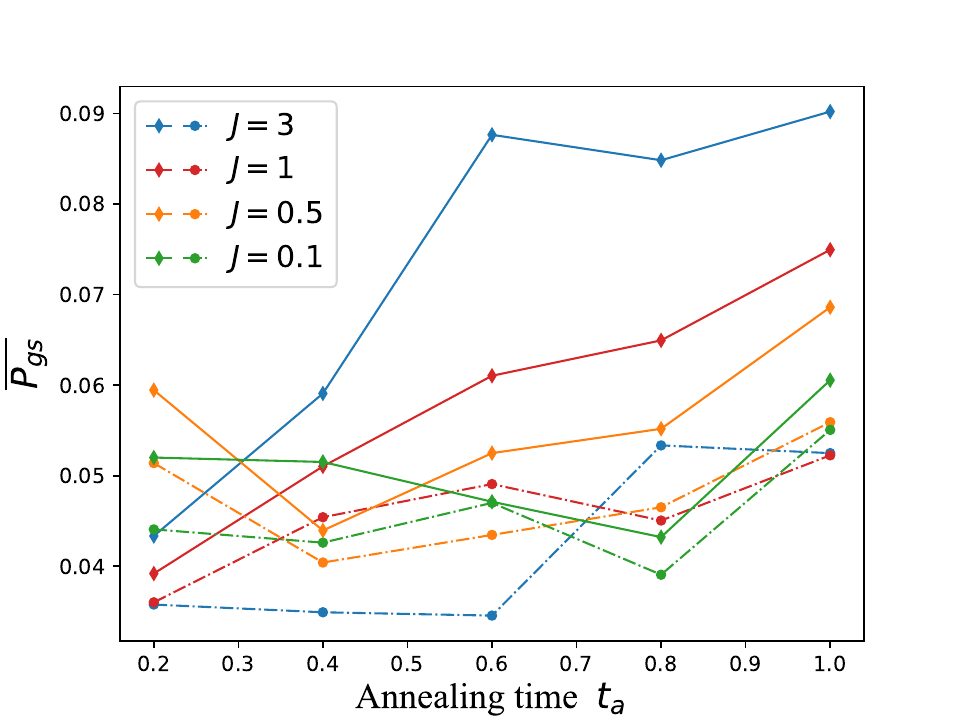}
\caption{Average ground state probabilities for a 5-qubit quantum chain on ibmq\_santiago. Solid lines are the catalyzed QAA results, dashed lines are the uncatalyzed QAA results. The catalyst Hamiltonian improves the probability substantially for $J = 0.5, 1, 3$. }
\label{fig:2}
\end{figure}

\section{Conclusions and Future Directions}\label{Conclusion}

We have demonstrated that inducing MBD by the introduction of the appropriate catalytic term can greatly speed up the QAA.  The speedups are clear simple calculation of the time required to reach a given speedup probability.  To make the specific connection to the MBL-MBD transition, we offered five pieces of evidence.  First, the fact that the speedup occurs at the value of the interaction parameter that corresponds to the known value at the transition.  Second, that the speedup is greatest that the time in the algorithm when the delocalization term is relatively the largest.  Third, that the speedup comes from increase in the gap rather than suppressing transition matrix elements.  Fourth, that the participation ratio, the classic signature of localization, is greatly increased by the delocalization term.  Finally, the entanglement entropy is strongly enhanced by the delocalization catalysis.  Thus, there is little doubt that we have identified the physical mechanism behind the speedup.

A great deal is known about low-dimensional spin systems in condensed matter physics. They show particularly strong quantum effects.  The method described here is an attempt to exploit these effects for the purpose of improving quantum adiabatic optimization.  Our simulations are necessarily limited to quite small systems, but the results are extremely encouraging. In particular, there are indications from scaling arguments that the speedups are not limited to small systems.

It is evident that the cost of the catalyzed and uncatalyzed algorithms are related by a constant factor on any gate-based machine.  That factor is of course important for a rigorous comparison of the two methods if the algorithmic speedup is itself a constant.  Our scaling results suggest but do not prove otherwise.  Such factors are also machine-dependent.  For a quantum annealing device, the factor may be close to unity and in that case these questions do not arise.

We have investigated only the RFIM problem.  The necessary quantum effects will likely be useful for other models that can be delocalized. It is possible it will work only on models defined on relatively sparse graphs.  When the coordination number increases, the spins become effectively more classical.  However, even in two dimensions we still found a speedup.  

It is important to point out that the concepts are not limited to the RFIM.  The method is based on identifying the paths that the system needs to follow in order to optimize its configuration efficiently.  In the RFIM, this is domain wall motion.  Then one chooses a catalyst that delocalizes the degrees of freedom for these paths.  We conjecture that if a catalyst Hamiltonian can be found that changes the model from one that is MBL to one that is MBD, then gap amplification and speedup of the QAA can be achieved, even in the degree of freedom is not a domain wall.  

In future work, one may reverse the logic to use the speedup in condensed-matter research to investigate the existence of the MBL-MBD transition in specific models.  

Finally, we note that the quantum optimization results may be improved by increasing the number of parameters and classically optimizing over different catalyst terms and the schedule of the catalysis.

\acknowledgments
We thank B. Ozguler, M. G. Vavilov, Tao Xiang and Yunlong Yu for useful discussions. We acknowledge the use of IBM Quantum services for this work.

\bibliography{ref}

\end{document}